\documentclass[twocolumn]{aastex63}
\begin{document}

\newcommand{\kms}{km~s$^{-1}$}	\newcommand{\cms}{cm~s$^{-2}$}
\newcommand{\msun}{$M_{\odot}$} \newcommand{\rsun}{$R_{\odot}$} 
\newcommand{\teff}{$T_{\rm eff}$} \newcommand{\logg}{$\log{g}$} 
\newcommand{\mas}{mas~yr$^{-1}$} 


\title{An Isolated White Dwarf with a 70 second Spin Period}

\author[0000-0001-6098-2235]{Mukremin Kilic} \affiliation{Homer L. Dodge Department 
of Physics and Astronomy, University of Oklahoma, 440 W. Brooks St., Norman, OK, 
73019 USA}

\author[0000-0002-9878-1647]{Alekzander Kosakowski} \affiliation{Department of Physics
and Astronomy, Texas Tech University, Lubbock, TX 79409, USA}

\author[0000-0001-7143-0890]{Adam G. Moss} \affiliation{Homer L. Dodge Department of Physics and Astronomy,
University of Oklahoma, 440 W. Brooks St., Norman, OK, 73019 USA}

\author[0000-0003-2368-345X]{P. Bergeron} \affiliation{D{\'e}partement de 
Physique, Universit{\'e} de Montr{\'e}al, C.P. 6128, Succ. Centre-Ville, 
Montr{\'e}al, Quebec H3C 3J7, Canada}

\author[0000-0002-2545-3597]{Annamarie A. Conly} \affiliation{Homer L. Dodge Department of Physics and Astronomy,
University of Oklahoma, 440 W. Brooks St., Norman, OK, 73019 USA}

\shortauthors{Kilic et al.}
\shorttitle{The Fastest Spinning Isolated WD}

\begin{abstract}

We report the discovery of an isolated white dwarf with a spin period of 70 s. We obtained high speed photometry of
three ultramassive white dwarfs within 100 pc, and discovered significant variability in one. SDSS
J221141.80+113604.4 is a $1.27~M_{\odot}$ (assuming a CO core) magnetic white dwarf that shows 2.9\%
brightness variations in the BG40 filter with a $70.32 \pm 0.04$ s period, becoming the fastest spinning
isolated white dwarf currently known. A detailed model atmosphere analysis shows that it has a mixed hydrogen and helium
atmosphere with a dipole field strength of $B_d = 15$ MG. Given its large mass, fast rotation, strong magnetic field,
unusual atmospheric composition, and relatively large tangential velocity for its cooling age, J2211+1136 displays all of
the signatures of a double white dwarf merger remnant. Long term monitoring of the spin evolution of J2211+1136 and
other fast spinning isolated white dwarfs opens a new discovery space for substellar and planetary mass companions
around white dwarfs. In addition, the discovery of such fast rotators outside of the ZZ Ceti instability strip suggests that some
should also exist within the strip. Hence, some of the mono-periodic variables found within the instability strip
may be fast spinning white dwarfs impersonating ZZ Ceti pulsators. 

\end{abstract}

\keywords{
        Magnetic variable stars ---
	White dwarf stars ---
	Compact objects ---
	Stellar remnants ---
	Periodic variable stars ---
	Short period variable stars
}

\section{INTRODUCTION}

Angular momentum transport between the core and the envelope should slow the rotation of the core during the giant
branch evolution, leading to the formation of slowly rotating white dwarfs \citep{kawaler04,tayar13}.  Measurement of
the white dwarf rotation rate is possible through high cadence observations of pulsating or spotted white dwarfs, or high
resolution spectroscopy of the NLTE H$\alpha$ core in DA white dwarfs. The latter indicates no or very slow rotation in
the majority of the observed systems, with typical lower limits of hours or longer rotation periods \citep{koester98}. 

Asteroseismology of isolated pulsating white dwarfs confirm these findings; average mass white dwarfs with
$M = $ 0.51-0.73 $M_{\odot}$ have a mean rotation period of $35 \pm 28$ h \citep{kawaler15,hermes17b}. The latter authors
discuss a possible link between white dwarf mass and rotation rate. There are three massive pulsating white dwarfs in their
sample with $M =$ 0.78-0.88 $M_{\odot}$, and those show significantly faster rotation rates of 1.1 to 8.9 h
\citep[see also][]{hermes17a}. 

Magnetic white dwarfs tend to spin faster than their non-magnetic counterparts. \citet{brinkworth13} found photometric
variability in 67\% of the isolated magnetic white dwarfs in their sample, with periods ranging from 27 minutes to 6 days,
plus two additional longer period systems. They also found no correlation between spin period and any other white dwarf
parameters, including mass. 

\citet{kawka20} and \citet{ferrario20} presented a summary of the rotation period
measurements for magnetic white dwarfs: most have rotation periods
shorter than 10 h, with a distribution that peaks at 2-3 h. Some of the fastest rotators are hot DQ white dwarfs with rotation periods as short as 5 min \citep{montgomery08,dufour11,williams16}. The combination of fast rotation rates, unique chemical composition,
high mass, and high incidence of magnetism in hot DQ white dwarfs favor a double white dwarf merger origin for the formation
of these stars \citep{dunlap15,coutu19}.

\begin{deluxetable*}{ccccc}
\tablecolumns{5}
\tablewidth{0pt}
\tablecaption{Details of APO 3.5m Agile Observations. \label{tab:obs}}
\tablehead{ \colhead{Object} & \colhead{Gaia DR2 Source ID} & \colhead{g (mag)} & \colhead{Exposures} & \colhead{UT Date}}
\startdata
SDSS J221141.80+113604.5   &  2727596187657230592 & 19.28 & $145 \times 30$ s, $281\times15$ s & 2021 Aug 5\\
\nodata & \nodata & \nodata & $720\times10$ s & 2021 Sep 9\\
SDSS J225513.48+071000.9   &  2712093451662656256 & 19.15 & $181 \times 15$ s & 2021 Aug 5\\
WD J010338.56$-$052251.96  &  2524879812959998592 & 17.41 & $1440 \times 5$ s & 2021 Sep 9\\
\enddata
\end{deluxetable*}

\begin{figure*} 
\vspace{-0.3in}
\hspace{-0.2in}
\includegraphics[width=2.5in, clip=true, trim=0.3in 2in 0.6in 1.4in]{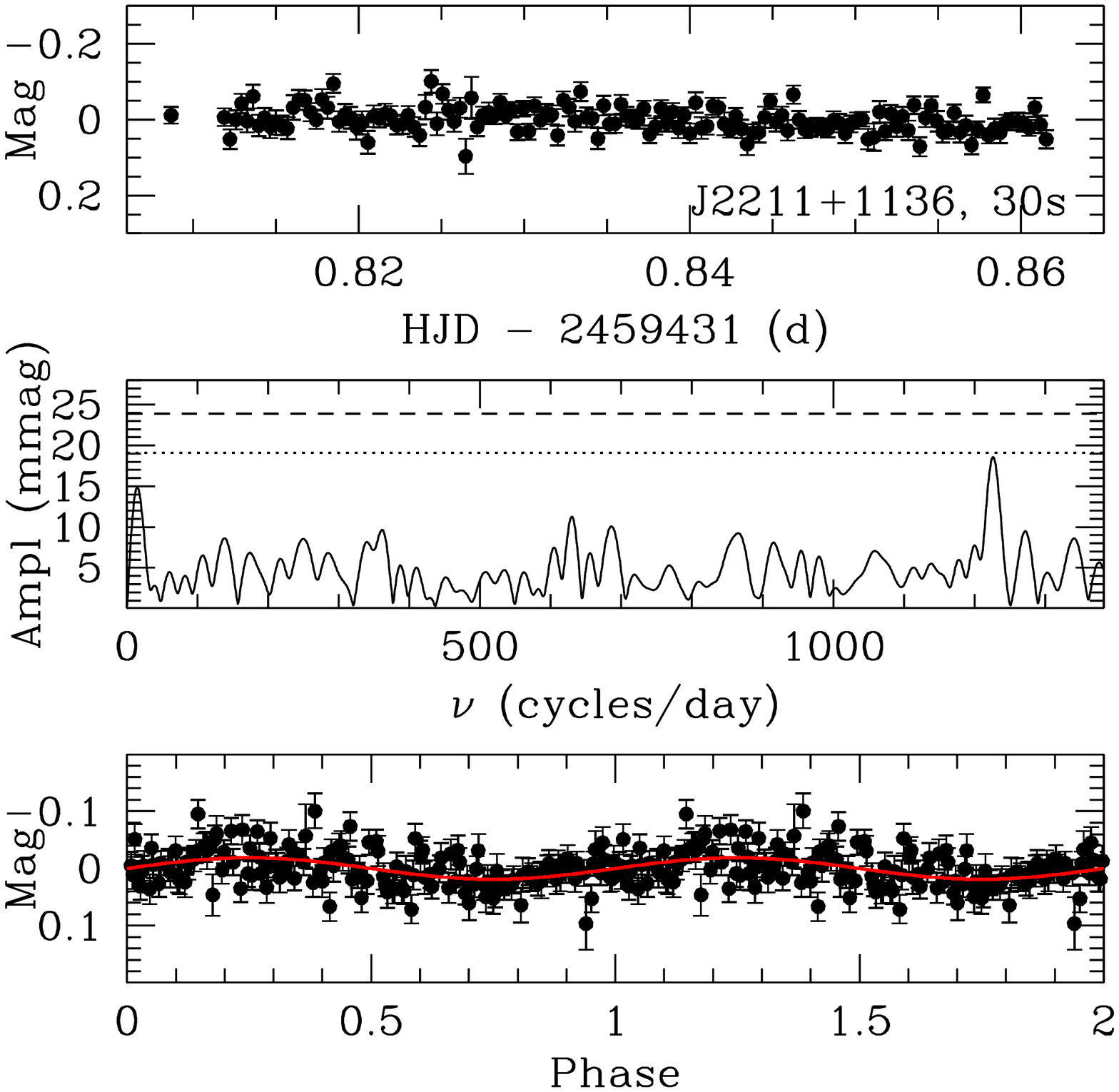}
\includegraphics[width=2.5in, clip=true, trim=0.3in 2in 0.6in 1.4in]{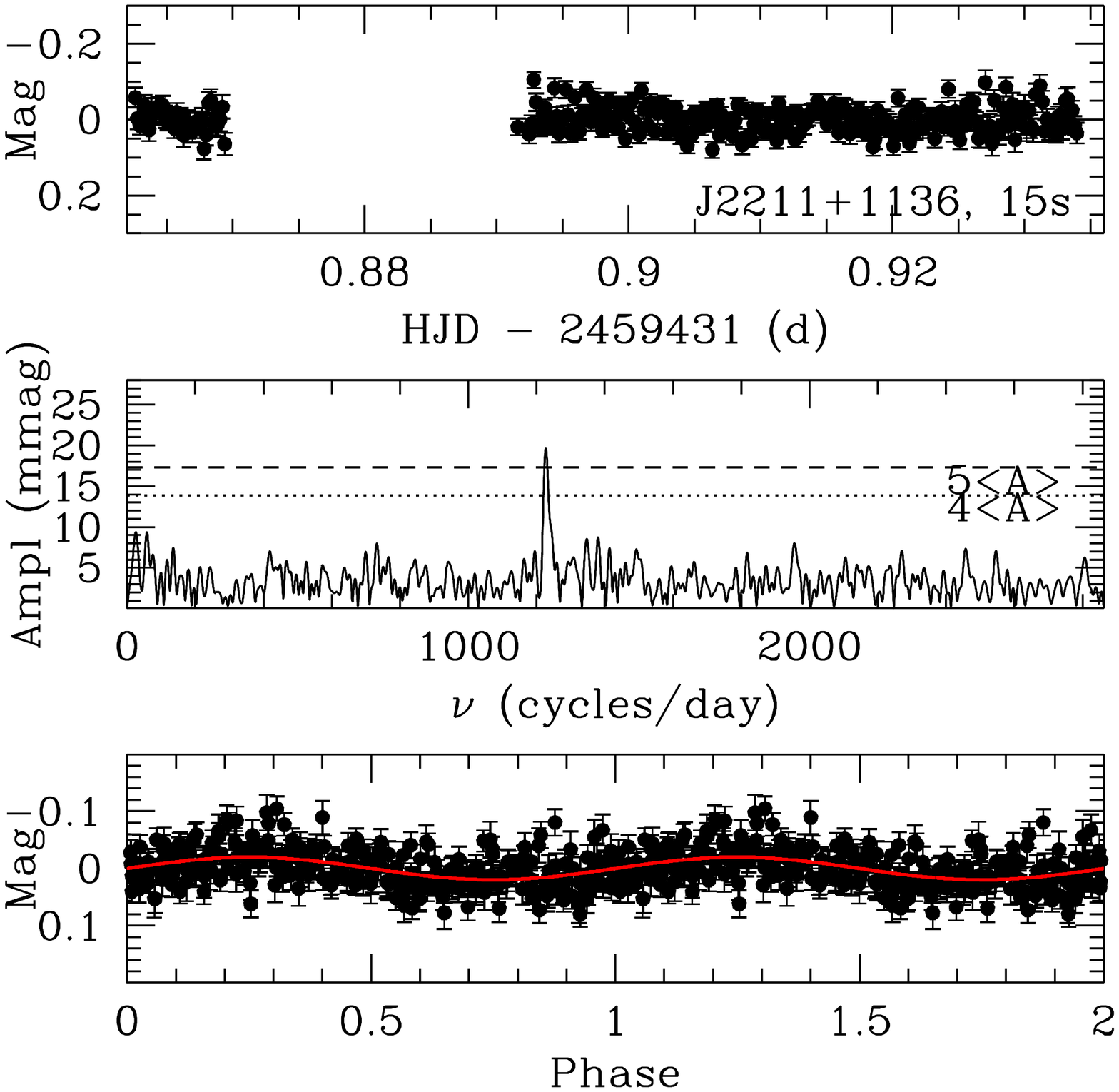}
\includegraphics[width=2.5in, clip=true, trim=0.3in 2in 0.6in 1.4in]{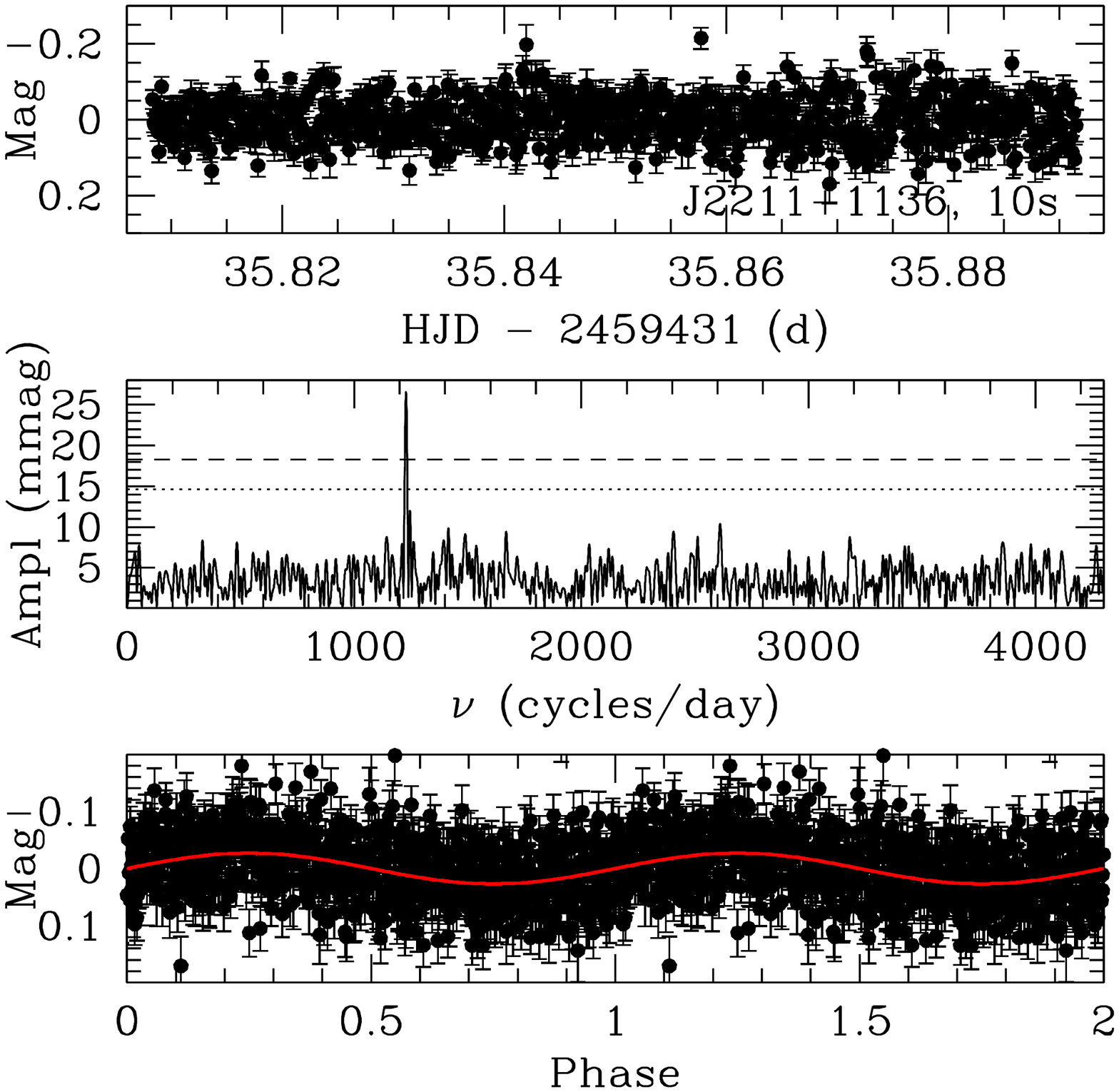}
\caption{{\it Top panels:} APO time-series photometry of J2211+1136 based on 30 (left), 15 (middle), and 10 s long exposures (right).  
{\it Middle panels:}  Fourier transform of each light curve. The dotted and dashed lines show the 4$\langle {\rm A}\rangle$ and
5$\langle {\rm A}\rangle$ levels, where $\langle {\rm A}\rangle$ is the average amplitude in the Fourier transform. {\it Bottom panels:}
The same light curves folded at the highest peak in the Fourier transform, along with the best-fitting sinusoidal model (red line).}
\label{fig2211}
\end{figure*}

\citet{kilic21} presented an analysis of the ultramassive ($M\geq1.3 M_{\odot}$) white dwarf candidates in the Montreal
White Dwarf Database \citep[MWDD,][]{dufour17} 100 pc sample, and identified four outliers in transverse velocity, four likely magnetic white dwarfs
(one of which is also an outlier in transverse velocity), and one with rapid rotation. They concluded that at least 32\% of
the 25 ultramassive white dwarfs in that sample are likely double white dwarf merger products. Among these ultramassive
white dwarfs, J183202.83+085636.24 was previously identified as a rapid rotator with a spin period of 353 s \citep{pshirkov20}.
Recently, \citet{caiazzo21} found a rotation period of 6.94 min in another of these objects, J190132.9+145808.7, based on
photometric variability detected in the Zwicky Transient Facility \citep{ztf}. These rotation rates are consistent with
the theoretical predictions for single white dwarfs that formed from double white dwarf mergers \citep{schwab21}.

There remains three additional confirmed or suspected magnetic white dwarfs known in the \citet{kilic21} ultramassive
white dwarf sample; SDSS J221141.80+113604.5 is a DAH with weak H$\alpha$ and H$\beta$ features, SDSS
J225513.48+071000.9 has a DC-like spectrum that shows broad unidentified features, and WD
J010338.56$-$052251.96 (G270-126) is a DAH: \citep{tremblay20}. We refer to these systems as J2211+1136, J2255+0710, and
J0103$-$0522, respectively. We obtained follow-up high speed time series photometry of these three systems to constrain
their rotation rates. Table \ref{tab:obs} presents the details of our observations for each target. We present the light curves
for J2211+1136 in Section 2, J2255+0710 and J0103$-$0522 in Section 3, discuss the variability in J2211+1136
and its implications in Section 4, and conclude.

\section{J2211+1136}

\subsection{Photometric Variability}

We acquired high speed photometry of J2211+1136 over 76 minutes on UT 2021 Aug 5 using the APO 3.5m telescope
with the Agile frame transfer camera and the BG40 filter. We obtained 30 s long back-to-back exposures
and binned the CCD by $2\times2$, which resulted in a plate scale of 0.258 arcsec pixel$^{-1}$. 

A quick reduction of these
data soon after acquisition revealed a frequency peak near 70 s, which was barely resolved due to our 30 s
long exposures. To improve our time resolution, we decreased the exposure time to 15 s, and obtained an additional set of
281 exposures on the same night. Our efforts to re-observe J2211+1136 on Aug 13 and Sep 1 failed due to the monsoon season,
but we were finally able to obtain 2 hours of 10 s long back-to-back exposures on UT 2021 Sep 9 under clear skies and
sub-arcsecond seeing.

Figure \ref{fig2211} shows these light curves based on 30 (left), 15 (middle), and 10 s (right) long exposures. The middle
panels show the Fourier transform of each dataset. All three datasets show a single peak around 70 s in
Fourier space, which is detected at the $\langle {\rm 4A}\rangle$ level in the 30 s cadence data, and
above the $\langle {\rm 5A}\rangle$ level in the higher cadence data. The bottom panels show each light curve
folded at the highest peak in the Fourier transform, along with the best-fitting sinusoidal model (red line).

Treating the 30, 15, and 10 s datasets separately, the former indicates $19\pm3$ millimag
variations with $P = 70.45 \pm 0.12$ s, whereas the second dataset indicates $20\pm3$ millimag variations with $P = 70.39 \pm 0.06$ s.
The higher cadence Sep 9 dataset (right panels) shows a larger peak in its Fourier transform at
$\nu = 1228.7 \pm 0.7$ cycles d$^{-1}$, or $P= 70.32 \pm 0.04$ s, with a $27 \pm 3$ millimag amplitude.
The period estimates from these three different datasets with 30, 15, and 10 s exposures agree within the errors.
Since the latter dataset has the longest baseline and highest cadence, it provides the best constraints on the period
and amplitude of the observed variations. Even then, because our 10 s exposures span 14.2\% of the rotation phase
($\delta\phi = 0.89$ radians), the observed amplitude is underestimated by a factor of $\sin{\delta\phi}/{\delta\phi} = 0.87$.
The true amplitude of variability is thus 31 millimag, or 2.9\%. 

\subsection{Model Atmosphere Analysis}

We computed magnetic model spectra using an approach similar to that described in \citet{bergeron92}, where the line
displacements and strengths of the Zeeman components are taken from the tables of \citet{kemic74}, with the exception
that here we include H$\alpha$ through H$\delta$. In addition to resonance broadening by neutral
hydrogen, and van der Waals broadening by neutral hydrogen and helium (included following the prescription of
\citealt{bergeron97}), we also take into account Stark broadening, which dominates the broadening of higher Balmer lines
at the temperature of the object studied here. The total line opacity can be expressed as the sum of the individual
Stark-, resonance-, and van der Waals-broadened Zeeman components. 

The specific intensities at the surface,
$I(\nu,\mu,\tau_{\nu}=0)$, are obtained by solving the radiative transfer equation for various field strengths and values
of $\mu$ ($\mu= \cos \theta$, where $\theta$ is the angle between the angle of propagation of light and the normal to the
surface of the star). Finally, the emergent spectrum is obtained from an integration over the surface of the star
($H_{\nu} \propto \int I_{\nu} \mu\,d\mu$) for a particular geometry of the magnetic field distribution. Here we
consider the same offset dipole model described in \citet{bergeron92}, where the independent parameters are the
dipole field strength $B_d$, the dipole offset $a_z$ measured in units of stellar radius from the center of the star,
and the viewing angle $i$ between the dipole axis and the line of sight ($i=0^{\circ}$ for a pole-on view).

\begin{figure}
\hspace{-0.2in}
\includegraphics[width=3.5in, clip=true, trim=0in 1.9in 0.4in 2.1in]{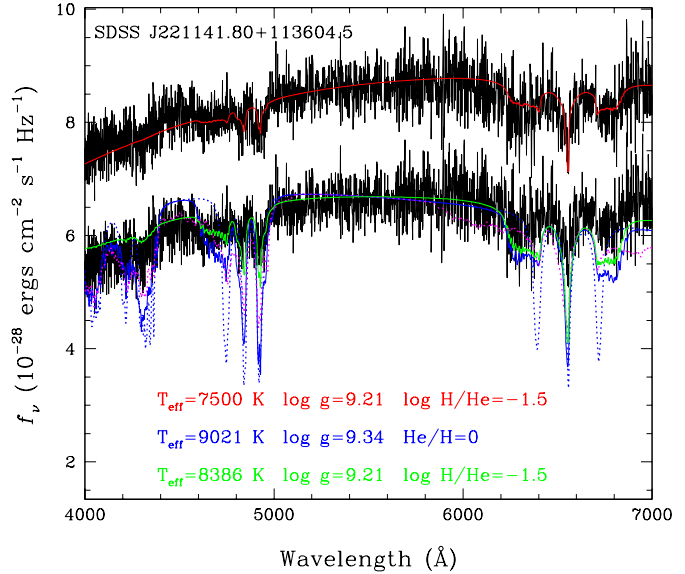}
\caption{Our best fits to the SDSS spectrum of the magnetic DA white dwarf J2211+1136. The lower fits
are for synthetic spectra calculated at temperatures and surface gravities determined from photometric fits under the
assumption of a pure hydrogen composition (solid blue) and a mixed composition of $\log {\rm H/He}=-1.5$ (solid green),
a dipole field strength of $B_d=15$ MG, a centered dipole ($a_z=0$), and a viewing angle of $i=45^{\circ}$.
All spectra are normalized at 5500 \AA. For comparison, we also show the offset dipole models for the pure H solution with
$a_z=-0.2$ and +0.2 as dotted blue (with very sharp Zeeman components) and magenta lines, respectively. 
The optical spectrum is reproduced at the top of the figure, arbitrarily shifted vertically for clarity,
and compared with a model spectrum where the effective temperature for the mixed H/He solution is decreased to
$T_{\rm eff}=7500$~K.\label{figx}}
\end{figure}

\begin{figure*} 
\includegraphics[width=3.5in, clip=true, trim=0.2in 0.2in 0.4in 0.6in]{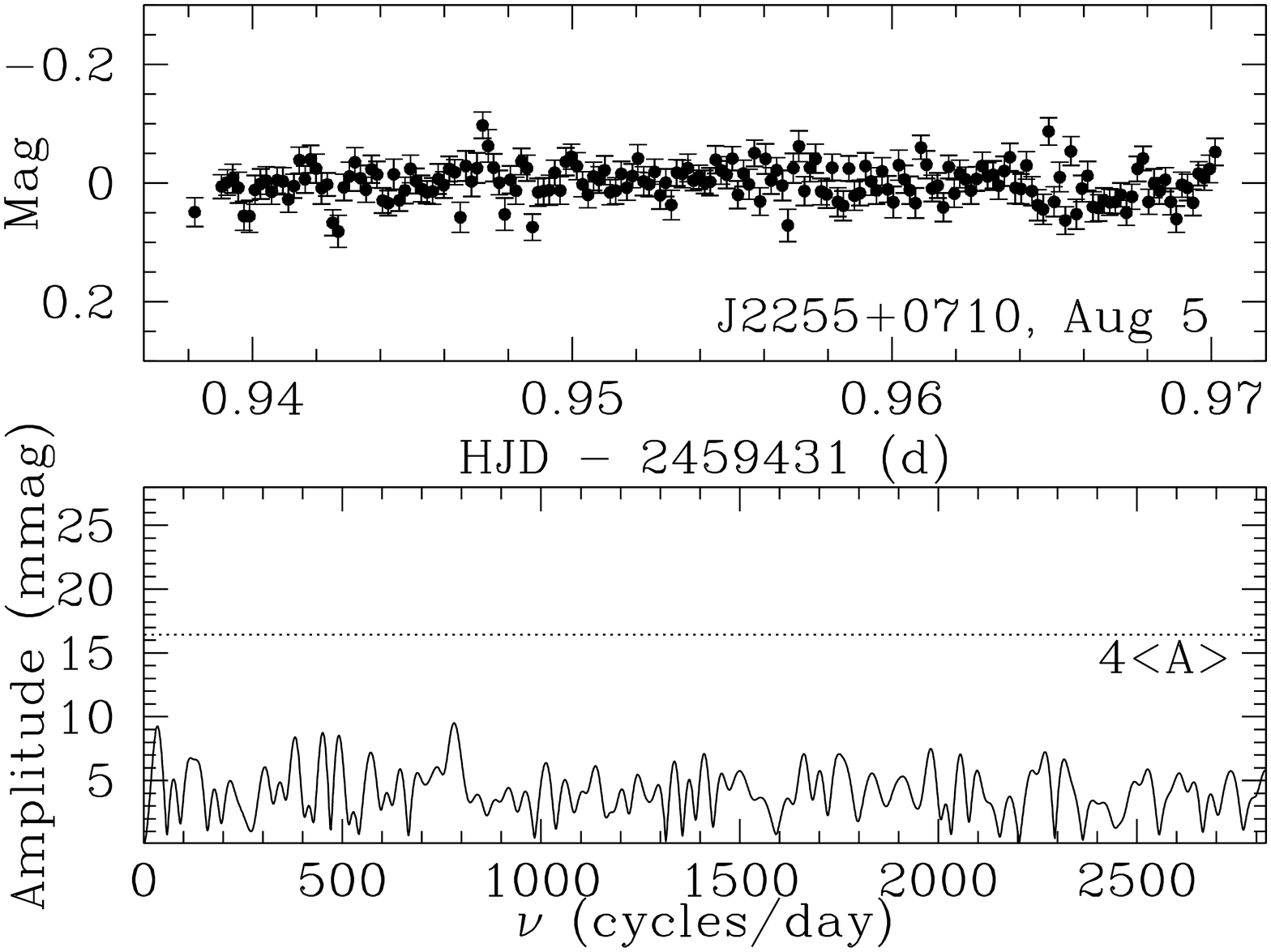}
\includegraphics[width=3.5in, clip=true, trim=0.2in 0.2in 0.4in 0.6in]{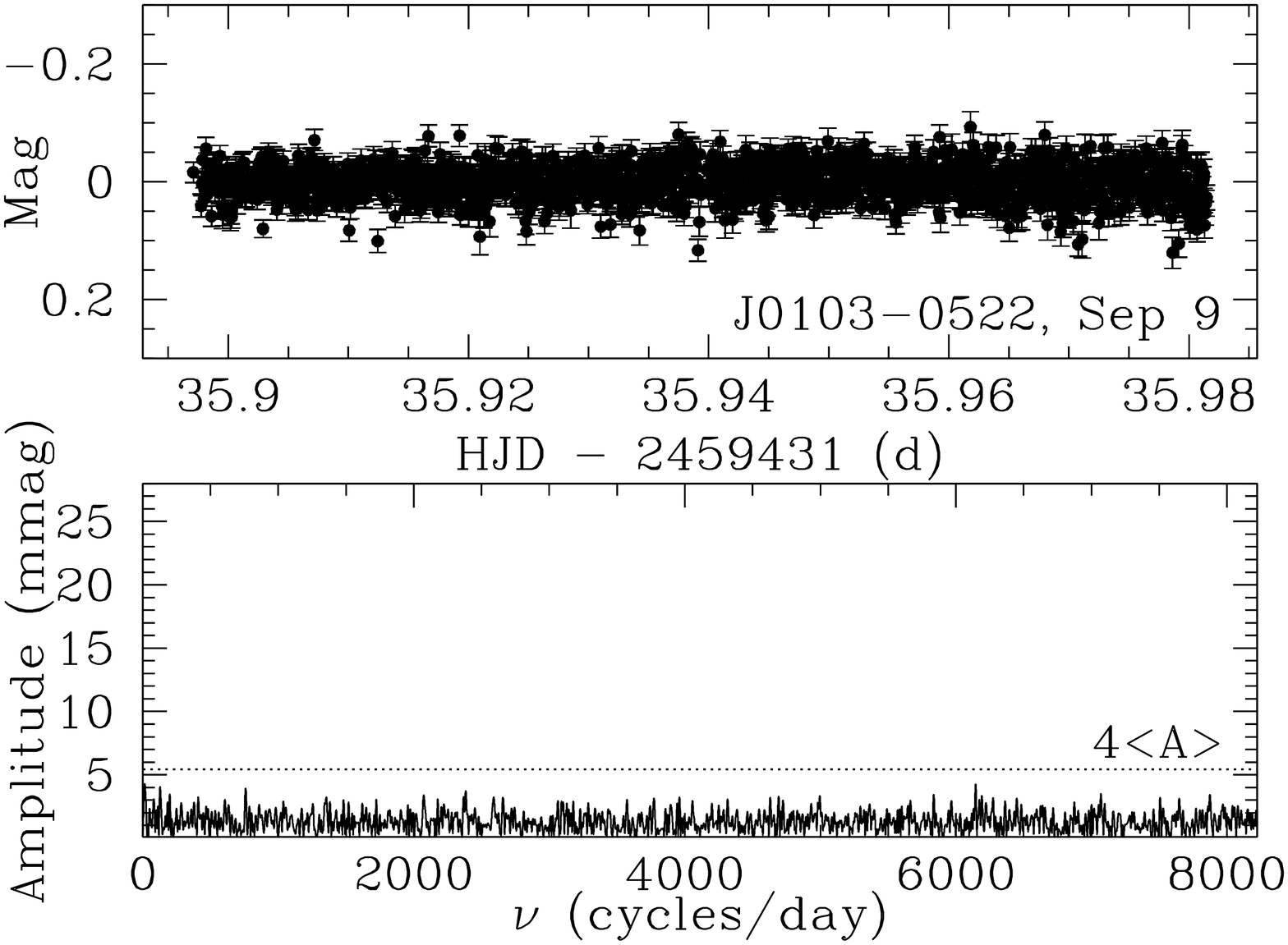}
\caption{APO light curves of J2255+0710 (top left) and J0103$-$0522 (top right) obtained on UT 2021 Aug 5 and Sep 9, respectively.
The bottom panels show the Fourier transform of these light curves, and the 3$\langle {\rm A}\rangle$ and
4$\langle {\rm A}\rangle$ levels.}
\label{fignon}
\end{figure*}

We first assume that J2211+1136 has a pure hydrogen composition, and measure
its effective temperature and stellar radius by fitting the available SDSS $u$ and Pan-STARRS $grizy$ photometry along
with the Gaia EDR3 parallax. Details of our fitting procedure, model atmosphere grid (including models with mixed H/He
compositions), as well as the evolutionary models used to derive the stellar mass and surface gravity, are described
in \citet{kilic20} and references therein. The best-fitting model under the assumption of a pure hydrogen composition
has $T_{\rm eff}=9021\pm 160$~K, $M=1.312\pm0.010$~\msun, and $\log{g}=9.338\pm 0.030$. However, a pure
hydrogen composition is clearly ruled out, since the higher Balmer lines are predicted way too strong, no matter
the assumed field strength and geometry. Figure \ref{figx} shows our magnetic model fits to the SDSS spectrum of
J2211+1136, including the pure hydrogen atmosphere solution (solid blue line). Here we simply assumed a dipole field
strength of $B_d=15$ MG, a centered dipole ($a_z=0$), and a viewing angle of $i=45^{\circ}$, which are the values
obtained for our best fit with a mixed H/He composition discussed in the next paragraph. For comparison, we also
show the offset dipole models with $a_z=-0.2$ and +0.2 as dotted blue and magenta lines, respectively. Even though
offset dipole models are commonly used in modeling magnetic white dwarfs \citep[e.g.,][]{rolland15}, a centered
dipole clearly provides a better fit to the observed line profiles in J2211+1136.

One obvious way to reduce the strength of the higher order Balmer lines is to assume that the star has a mixed H and He
composition (see, e.g., Figure 13 of \citealt{bergeron91}). Such He-rich DA stars have been identified in large numbers
in the SDSS \citep{rolland18}.  We therefore refitted the photometric energy distribution, but this time by assuming various
values of the hydrogen-to-helium abundance ratio in number, H/He. We then explored for each solution different values of
the field strength and offset, and considered two viewing angles $i=45^{\circ}$ and $60^{\circ}$. We also explored rotational
broadening and found that it only affects the line core. Our best overall fit, shown
in green in Figure \ref{figx}, is achieved with a mixed composition of $\log {\rm H/He}=-1.5$, from which we measure
$T_{\rm eff}=8386\pm 267$~K, $M=1.268\pm0.010$~\msun, and $\log{g}=9.214\pm 0.027$, for the same
field geometry as above ($B_d=15$ MG, $a_z=0$, and $i=45^{\circ}$).

Clearly, the mixed H/He solution provides a much better fit to the optical spectrum of J2211+1136 than
the pure hydrogen model. However, both the strength of H$\alpha$ and the slope of the energy distribution suggest a
lower temperature. We arbitrarily lowered the effective temperature of the mixed H/He solution to
$T_{\rm eff}=7500$~K, and found that this model provides an excellent fit to the optical spectrum. This solution is
displayed in red in Figure \ref{figx}. 

There are likely two reasons for the significant temperature difference between the photometric 
and spectroscopic solutions. First, the SDSS spectrum of J2211+1136 is a combination of eight
sub-exposures with a total exposure time of 7207 s; it covers $>100$ rotation periods 
 (see Section \ref{var}). Hence, any spectral changes due the magnetic field geometry or 
surface inhomogeneities would lead to additional smearing of the Zeeman split lines in the combined
SDSS spectrum. Second, \citet{kulebi09} noted that no atomic data for hydrogen in the presence
of both a magnetic and electric field are available for arbitrary strengths and arbitrary angles 
between the two fields. Therefore, there may be systematic uncertainties in the line profile 
calculations, which could lead to differences between the temperature estimates from the continuum 
slope and the line profiles \citep{kulebi09}. Regardless of these issues, we can safely conclude 
that J2211+1136 is a magnetic and mixed H/He atmosphere white dwarf with
$M=1.268\pm0.010$~\msun, $\log{g}=9.214\pm 0.027$, and $T_{\rm eff} \approx$ 7500 - 8390 K.

\begin{figure*} 
\center
\includegraphics[width=3.2in, clip=true, trim=0.1in 2in 0.3in 0.9in]{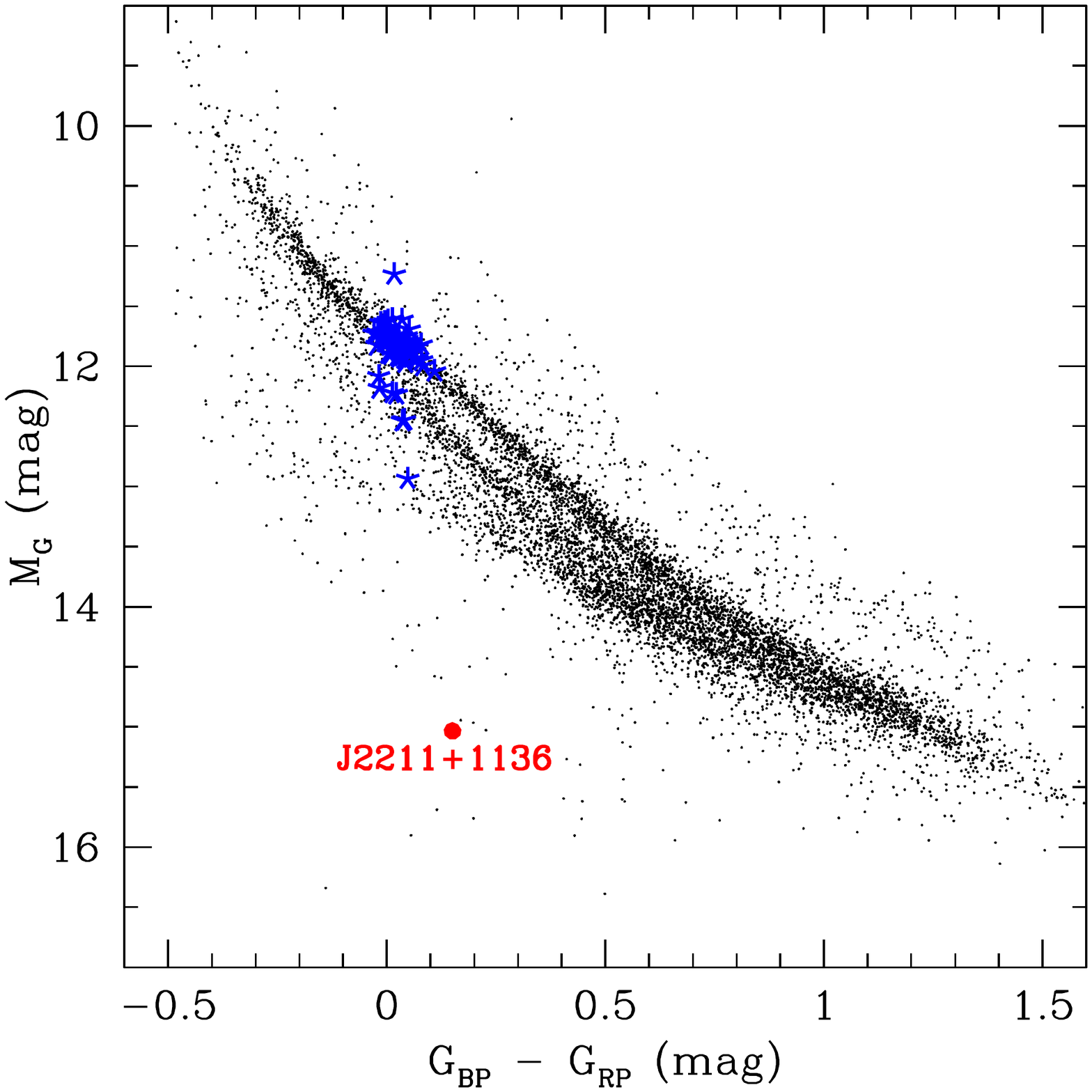}
\includegraphics[width=3.2in, clip=true, trim=0.1in 2in 0.3in 0.9in]{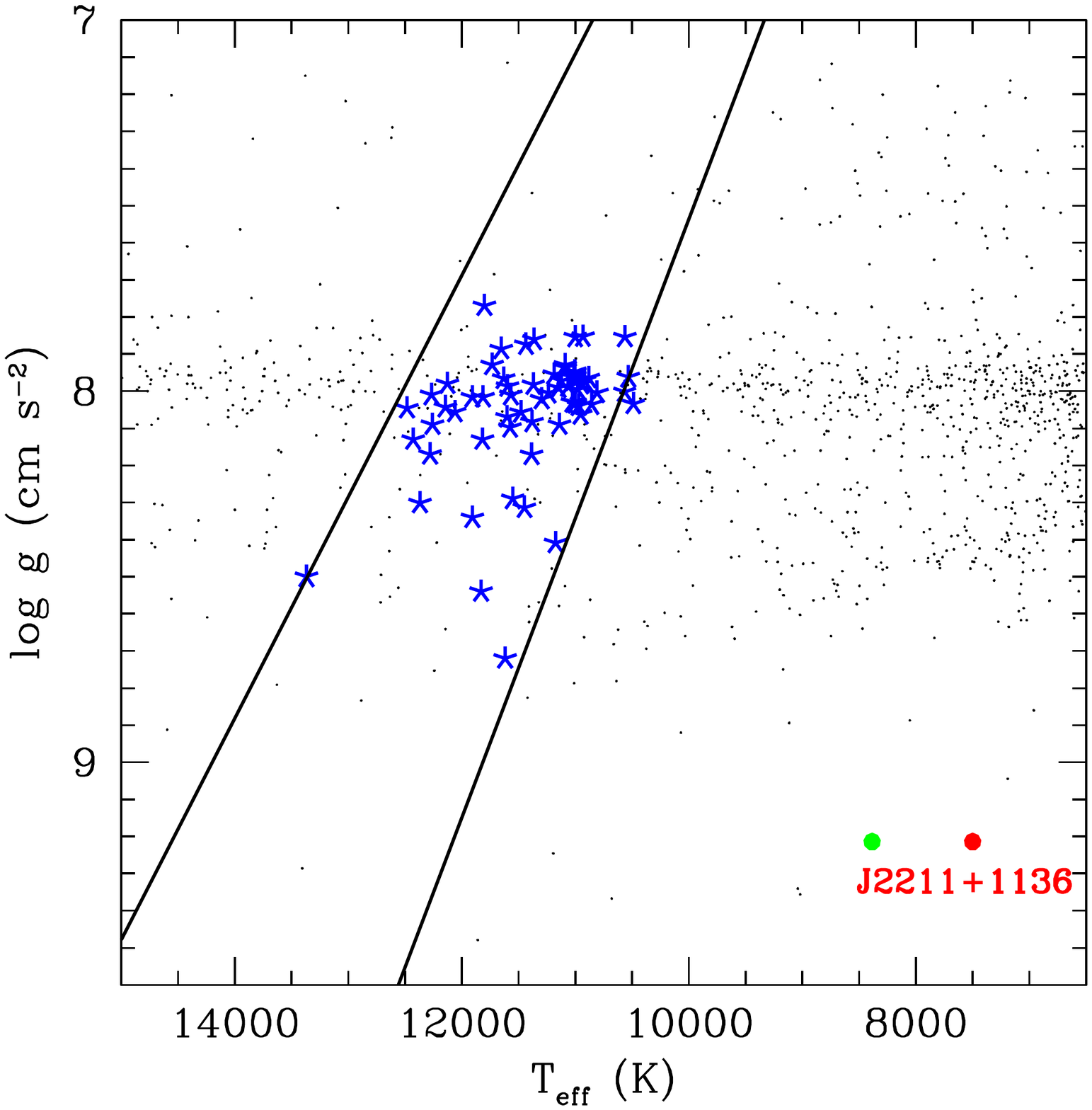}
\caption{{\it Left:} Gaia color-magnitude diagram of the 100 pc sample in the Montreal White Dwarf Database
\citep{dufour17}. {\it Right:} Temperatures and surface gravities of the same stars. Blue stars mark the previously known
pulsating DAV white dwarfs, and the solid lines mark the boundaries of the ZZ Ceti instability strip \citep{tremblay15}.
The atmospheric parameters of J2211+1136 based on the photometric (green) and the spectroscopic (red) method
are also shown. J2211+1136 is clearly outside the boundaries of the instability strip.}
\label{figstrip}
\end{figure*}

\section{J2255+0710 and J0103$-$0522}

We acquired high-speed photometry of J2255+0710 and J0103$-$0522 right after we observed J2211+1136 on Aug 5 and
Sep 9, respectively. We observed J2255+0710 over 45 min with 15 s long exposures, and J0103$-$0522 over  2 h with
5 s long exposures. Figure \ref{fignon} shows the APO light curves and Fourier transforms for both stars. Neither star
shows any significant variability, and we can rule out variability above 16.4 millimag in J2255+0710 and 5.4 millimag in
J0103$-$0522 at the 4$<$A$>$ level. Additional follow-up data on J2255+0710 would be useful in pushing this detection limit
down to lower amplitudes. 

Out of the four magnetic ultramassive white dwarfs discussed in \citet{kilic21}, two show photometric variability,
J2211+1136 (discussed here) and J1901+1458 \citep{caiazzo21}. This fraction, 50\%, is comparable to the
67\% fraction found in the larger variability survey of \citet{brinkworth13}. Even though J2255+0710 and J0103$-$0522
do not show large photometric variations, it is still possible that they could be fast rotators. For example, \citet{kilic19}
detected significant changes in the H$\alpha$ line profiles of G183-35 due to rotation, but G183-35 shows only 
low level photometric variability, at 0.2\%. Such a signal would be lost in the noise in our observations of
J2255+0710 and J0103$-$0522. 

\section{Discussion}

\subsection{The Source of Variability in J2211+1136}
\label{var}

There are two potential mechanisms to explain minute-scale variations in white dwarfs; pulsations and rotation.
Figure \ref{figstrip} shows the ZZ Ceti instability strip for DA white dwarfs in color-magnitude and
$T_{\rm eff}$-$\log{g}$ space using the 100 pc MWDD white dwarf sample \citep{dufour17}. Blue stars mark the
previously known pulsating DAV white dwarfs in that sample, and the solid lines show the empirical boundaries
of the instability strip \citep[see][and references therein]{tremblay15}. The atmospheric parameters of J2211+1136
based on the photometric (green) and the spectroscopic (red, see Fig. \ref{figx}) method are also shown.
J2211+1136 is significantly redder and cooler than the known pulsating DA white dwarfs, and it is clearly outside
the instability strip. The period of variability in J2211+1136 (70 s) is also significantly shorter than that in the shortest
period ZZ Ceti pulsators ($\approx$100 s).

Multiperiodicity is common among pulsating DAV white dwarfs \citep{mukadam04,fontaine08,winget08}. For example, BPM 37093 is
a $T_{\rm eff} = 11620 \pm 190$ K, $M=1.13 \pm 0.10~M_{\odot}$ \citep[assuming a CO core,][]{bedard17} pulsating
white dwarf that displays eight pulsation modes between 512 and 635 s \citep{metcalfe04}. Similarly, GD 518
is a $T_{\rm eff} = 11420 \pm 110$ K, $M=1.114 \pm 0.006~M_{\odot}$ \citep{kilic20} pulsating white dwarf that displays
multi-periodic luminosity variations at timescales ranging from about 425 to 595 s \citep{hermes13}. 

J2211+1136 is a He-rich DA white dwarf. It does not look like a ZZ Ceti star; it has a mixed H/He atmosphere and its surface
temperature is too cool for pulsations. It also does not sound like a ZZ Ceti, it only shows monoperiodic variability. Hence,
it certainly is not a pulsating white dwarf. The only other explanation for such short period variability is the rotation of a spotted
white dwarf.

\subsection{The fastest spinning isolated white dwarf}

Accreting white dwarfs in binary systems can be spun up to extremely fast rotation rates \citep[e.g.,][]{oliveira20},
RX J0648.0$-$4418 being an excellent example with a spin period of 13.2 s \citep{mereghetti21}. However, isolated
white dwarfs that go through single star evolution rotate relatively slowly \citep{kawaler15,hermes17b}.  

Modeling the evolution of double white dwarf merger remnants, \citet{schwab21} predict rotation periods as short
as 5min for the most massive white dwarfs with $M\gtrsim1.2~M_{\odot}$. They also find that increasing mass ratio
(at fixed total mass) leads to higher final masses and shorter periods, and conclude that a rotation period of
$\sim10$ min in a single white dwarf is a signature of its merger origin.

Prior to the discovery observations presented here, the fastest spinning isolated white dwarfs had rotation periods
of $\approx$ 5 min \citep{williams16,kawka20,pshirkov20,reding20,caiazzo21}. With a rotation rate of $P= 70.32 \pm 0.04$ s,
J2211+1136 becomes the fastest spinning isolated white dwarf known. In addition to its fast rotation, J2211+1136 is strongly
magnetic, massive, an outlier in its transverse velocity, and also an outlier in its atmospheric composition
\citep[see Fig. 8 in][]{rolland18}. Hence, J2211+1136 presents all of the symptoms of being a double white dwarf merger product. 
 
\subsection{Fast spinning white dwarfs impersonating ZZ Cetis}

Pure hydrogen atmosphere white dwarfs are expected to enter the ZZ Ceti instability strip once they cool down to about
12000 K \citep[for average $0.6~M_{\odot}$ white dwarfs,][]{fontaine08,winget08}. These non-radial $g$-mode pulsations have periods ranging from 100 s to 2000 s. The purity
of the ZZ Ceti instability strip has been discussed in length by \citet{mukadam04} and \citet{vincent20}. It is expected that
every pure hydrogen atmosphere white dwarf passing through the instability strip during their evolution should pulsate. However,
the presence of a magnetic field may significantly affect the pulsation driving mechanism \citep{tremblay15}. The case of
WD 2105$-$820 is especially interesting; a $\sim56$ kG field seems to be sufficient to
suppress atmospheric convection \citep{gentile18}, and therefore inhibit pulsations. This suggests that magnetic white dwarfs should
not pulsate, even if they are within the confines of the instability strip \citep{vincent20}. 

The presence of magnetic white dwarfs within the instability strip implies that some of them will show mono-periodic photometric
variations due to rotation. Such fast rotating magnetic white dwarfs can impersonate a pulsating ZZ Ceti star. 
For example, \citet{kilic15} discovered photometric variations in a massive DA white dwarf that is very close to the red edge
of the instability strip, J1529+2928. However, given the relatively long period of 38 min, they ruled out pulsations as the
source of variability, and instead suggested a spotted white dwarf. If the rotation period of J1529+2928 was less than about 20 min,
it would have been classified as a ZZ Ceti. \citet{curd17} identified 4 massive pulsating DA white dwarfs, but three of them
show mono-periodic variability, even though mono-periodic pulsators are rare \citep[e.g.,][]{mukadam04,hermes17b}.
Unless higher signal-to-noise ratio observations detect additional pulsation modes, it is difficult to confirm these mono-periodic
variables as pulsating DAVs.
     
\section{Conclusions and Future Directions}

We presented follow-up time-series photometry of three ultramassive white dwarfs in the 100 pc sample \citep{kilic21},
and reported the discovery of 70.3 s photometric variations in one of these systems. J2211+1136 becomes the fastest
spinning isolated white dwarf known. J2211+1136 shows all of the signatures of a binary merger outcome; it is strongly
magnetic, fast rotator, ultramassive, and has a relatively large transverse velocity and an unusual atmospheric composition. 

\citet{briggs15} suggest that binary mergers can explain the incidence of magnetism and the mass distribution of highly
magnetic white dwarfs. Since mergers also lead to fast rotation rates \citep{schwab21}, it is natural to expect a trend
in mass and rotation. So far, only a small fraction of the ultramassive white dwarfs in the solar neighborhood have
spectral classification available. Additional  follow-up spectroscopy and high speed photometry would be useful to
search for additional magnetic white dwarfs and fast rotators, and search for trends between the rotation rates and
other physical parameters of these objects \citep{brinkworth13,hermes17b}. 

Large scale photometric surveys like the ZTF \citep{ztf} and the Vera Rubin Observatory's Legacy Survey
of Space and Time (LSST) will provide an unprecedented opportunity to enlarge the sample of spotted white dwarfs
with rotation measurements \citep[e.g.,][]{caiazzo21}. As an example, a quick search of the ZTF photometry for the
ultramassive white dwarfs presented in \citet{kilic21} shows significant variability for WD J070753.00+561200.25
with a 63 min period. Expanding this search to the (candidate) magnetic white dwarfs with $M\geq1~M_{\odot}$ in the 100 pc
SDSS sample of \citet{kilic20} reveals four additional variables, SDSS J011810.32$-$015612.3,
J071816.40+373139.0, J103941.52$-$032534.2, and J154315.09+302133.5. There are many other
fast rotating magnetic white dwarfs waiting to be discovered in the ZTF (I. Caiazzo 2021, private communication), and
eventually in the LSST. 

The existence of fast rotating white dwarfs outside of the ZZ Ceti strip indicates that there must be some within the strip.
Such mono-periodic variability can be easily confused with non-radial $g$-mode pulsations. Since a large fraction of massive
white dwarfs form through mergers \citep{temmink20} and such remnants are also likely to be fast spinning magnetic white
dwarfs \citep{briggs15,schwab21}, we would expect the fraction of ZZ Ceti impostors to be larger for more massive white dwarfs.
Separating the true pulsators from the ZZ Ceti impostors would require the detection of multi-periodic variations (for ZZ Cetis)
or a magnetic field (for the impostors) through high resolution spectroscopy or spectropolarimetry. 

Fast rotating white dwarfs with photometric variations provide an excellent clock to time the system. Stellar pulsations
have been used to search for planetary companions around white dwarfs \citep{winget03,mullally08}, but stable pulsation
modes are essential for a successful search \citep{hermes13b}. The timing method works best for stars with high amplitude,
short period oscillations that are stable. 

Timing measurements of millisecond pulsars identified the first planetary objects
outside of the solar system \citep{wolszczan94}. J2211+1136 and other fast spinning isolated white dwarfs with significant
photometric variability \citep[e.g.,][]{williams16,kawka20,pshirkov20,reding20,caiazzo21} open a new discovery space
for similar objects around white dwarfs. Long term monitoring of the spin evolution of these isolated white dwarfs can provide
meaningful constraints on the occurrence rate of substellar and planetary mass companions around their progenitor systems.

\acknowledgements

This work was supported in part by the NSF under grant AST-1906379, the NSERC Canada, and by the Fund FRQ-NT (Qu\'ebec).
The Apache Point Observatory 3.5-meter telescope is owned and operated by the Astrophysical Research Consortium.
\facilities{APO 3.5m (Agile)}


\end{document}